\documentclass{article}                     

\usepackage{apacite}

\usepackage{graphicx}

\usepackage{float}
\usepackage{multirow}
\usepackage{subfigure}
\usepackage{verbatim}
\usepackage[dvipsnames]{xcolor}
\usepackage[figuresright]{rotating}
\usepackage[normalem]{ulem}
\usepackage[title]{appendix}

\usepackage{url}
\usepackage{algorithmic}

\graphicspath{ {figures/} }

\usepackage[utf8]{inputenc}
\usepackage{amsmath,amssymb}
\usepackage{amsthm}

\usepackage{pdflscape}
\usepackage{afterpage}
\usepackage{capt-of}
\usepackage{fullpage}
\usepackage{tikz}

\usepackage[ruled,commentsnumbered,linesnumbered,vlined]{algorithm2e}

\usepackage{indentfirst}
\usepackage[normalem]{ulem}
\usepackage{textcomp}

\usepackage{wasysym} 
\usepackage{geometry}

\newtheorem{lemma}{Lemma}

\newtheorem{proposition}{Proposition}
\newcommand{\rafaelC}[1]{\textcolor{black}{#1}}

       \makeatletter
\def\@fnsymbol#1{\ensuremath{\ifcase#1\or *\or \mathsection\or \mathparagraph\or **\or \dagger\or \ddagger\or
    \|\or \dagger\dagger
   \or \ddagger\ddagger \else\@ctrerr\fi}}
    \makeatother

\title{New formulations and branch-and-cut procedures\\ for the longest induced path problem}

\author{
Rusl\'an G. Marzo\thanks{
Universidade Federal Fluminense, Institute of Computing, Niter\'oi, RJ 24210-240, Brazil. ({\tt ruslangm@id.uff.br})
}
\and 
Rafael A. Melo \thanks{Universidade Federal da Bahia, \rafaelC{Institute of Computing, Salvador, BA 40170-115,} Brazil. ({\tt \rafaelC{rafael.melo@ufba.br}}) 
}
\and Celso C. Ribeiro\thanks{
Universidade Federal Fluminense, Institute of Computing, Niter\'oi, RJ 24210-240, Brazil. ({\tt celso@ic.uff.br})
}
\and Marcio C. Santos {\thanks{Universidade Federal do Ceará, Campus Russas. Rua Felipe Santiago, 411. Russas, CE 62900-000. Brazil. ({\tt marciocs@ufc.br})}} 
}

\date{\today}

\begin{document}

\maketitle

\begin{abstract}

Given an undirected graph $G=(V,E)$, the longest induced path problem (LIPP) consists of obtaining a maximum cardinality subset $W\subseteq V$ such that $W$ induces a simple path in $G$. 
In this paper, we propose two new formulations with an exponential number of constraints for the problem, together with effective branch-and-cut procedures for its solution. While the first formulation (cec) is based on constraints that explicitly eliminate cycles, the second one (cut) ensures connectivity via cutset constraints.
We compare, both theoretically and experimentally, the newly proposed approaches with a state-of-the-art formulation recently proposed in the literature.
More specifically, we show that the polyhedra defined by formulation cut and that of the formulation available in the literature are the same. Besides, we show that these two formulations are stronger in theory than cec.
We also propose a new branch-and-cut procedure using the new formulations.
Computational experiments show that the newly proposed formulation cec, although less strong from a theoretical point of view, is the best performing approach as it can solve all but one of the 1065 benchmark instances used in the literature within the given time limit. In addition, our newly proposed approaches outperform the state-of-the-art formulation when it comes to the median times to solve the instances to optimality.
Furthermore, we perform extended computational experiments considering more challenging and hard-to-solve larger instances and evaluate the impacts on the results when offering initial feasible solutions (warm starts) to the formulations.
  \\


\end{abstract}

\setlength{\unitlength}{4144sp}%
\begingroup\makeatletter\ifx\SetFigFont\undefined%
\gdef\SetFigFont#1#2#3#4#5{%
  \reset@font\fontsize{#1}{#2pt}%
  \fontfamily{#3}\fontseries{#4}\fontshape{#5}%
  \selectfont}%
\fi\endgroup%

\section{Introduction}
\label{s_intro}

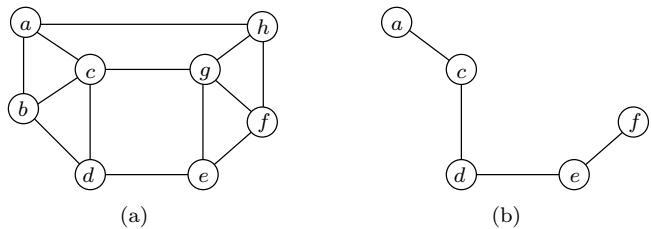
\begin{figure}[ht!]
    \begin{center}
      \subfigure[]{
                \begin{picture}(0,0)%
                \includegraphics{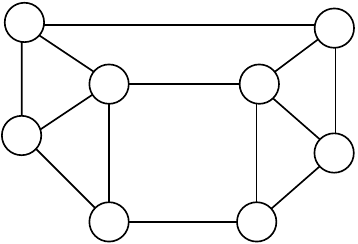}%
                \end{picture}%
                \setlength{\unitlength}{4144sp}%
                \begingroup\makeatletter\ifx\SetFigFont\undefined%
                \gdef\SetFigFont#1#2#3#4#5{%
                  \reset@font\fontsize{#1}{#2pt}%
                  \fontfamily{#3}\fontseries{#4}\fontshape{#5}%
                  \selectfont}%
                \fi\endgroup%
                \begin{picture}(1626,1106)(-92,-833)
                \put(361,-781){\makebox(0,0)[lb]{\smash{{\SetFigFont{8}{9.6}{\rmdefault}{\mddefault}{\updefault}{\color[rgb]{0,0,0}$d$}%
                }}}}
                \put(-14,131){\makebox(0,0)[lb]{\smash{{\SetFigFont{8}{9.6}{\rmdefault}{\mddefault}{\updefault}{\color[rgb]{0,0,0}$a$}%
                }}}}
                \put(-24,-387){\makebox(0,0)[lb]{\smash{{\SetFigFont{8}{9.6}{\rmdefault}{\mddefault}{\updefault}{\color[rgb]{0,0,0}$b$}%
                }}}}
                \put(377,-155){\makebox(0,0)[lb]{\smash{{\SetFigFont{8}{9.6}{\rmdefault}{\mddefault}{\updefault}{\color[rgb]{0,0,0}$c$}%
                }}}}
                \put(1060,-145){\makebox(0,0)[lb]{\smash{{\SetFigFont{8}{9.6}{\rmdefault}{\mddefault}{\updefault}{\color[rgb]{0,0,0}$g$}%
                }}}}
                \put(1406,-460){\makebox(0,0)[lb]{\smash{{\SetFigFont{8}{9.6}{\rmdefault}{\mddefault}{\updefault}{\color[rgb]{0,0,0}$f$}%
                }}}}
                \put(1054,-779){\makebox(0,0)[lb]{\smash{{\SetFigFont{8}{9.6}{\rmdefault}{\mddefault}{\updefault}{\color[rgb]{0,0,0}$e$}%
                }}}}
                \put(1403,108){\makebox(0,0)[lb]{\smash{{\SetFigFont{8}{9.6}{\rmdefault}{\mddefault}{\updefault}{\color[rgb]{0,0,0}$h$}%
                }}}}
                \end{picture}%

        } \hspace{1cm}
        	  \subfigure[]{
        	  
                        \begin{picture}(0,0)%
                    \includegraphics{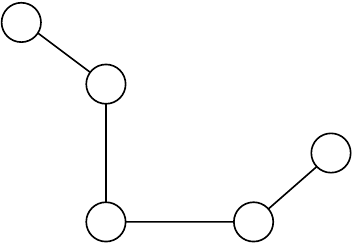}%
                    \end{picture}%
                    \setlength{\unitlength}{4144sp}%
                    \begingroup\makeatletter\ifx\SetFigFont\undefined%
                    \gdef\SetFigFont#1#2#3#4#5{%
                      \reset@font\fontsize{#1}{#2pt}%
                      \fontfamily{#3}\fontseries{#4}\fontshape{#5}%
                      \selectfont}%
                    \fi\endgroup%
                    \begin{picture}(1612,1106)(-79,-833)
                    \put(361,-781){\makebox(0,0)[lb]{\smash{{\SetFigFont{8}{9.6}{\rmdefault}{\mddefault}{\updefault}{\color[rgb]{0,0,0}$d$}%
                    }}}}
                    \put(-14,131){\makebox(0,0)[lb]{\smash{{\SetFigFont{8}{9.6}{\rmdefault}{\mddefault}{\updefault}{\color[rgb]{0,0,0}$a$}%
                    }}}}
                    \put(377,-155){\makebox(0,0)[lb]{\smash{{\SetFigFont{8}{9.6}{\rmdefault}{\mddefault}{\updefault}{\color[rgb]{0,0,0}$c$}%
                    }}}}
                    \put(1406,-460){\makebox(0,0)[lb]{\smash{{\SetFigFont{8}{9.6}{\rmdefault}{\mddefault}{\updefault}{\color[rgb]{0,0,0}$f$}%
                    }}}}
                    \put(1054,-779){\makebox(0,0)[lb]{\smash{{\SetFigFont{8}{9.6}{\rmdefault}{\mddefault}{\updefault}{\color[rgb]{0,0,0}$e$}%
                    }}}}
                    \end{picture}%
        	  
        }
    \end{center}
    \caption{Examples of (a) an input graph $G$ with node set $V=\{a,b,c,d,e,f,g,h\}$ and (b) a longest induced path $G[W]$ with $W=\{a,c,d,e,f\}$.}\label{fig:longestinducedpath}
\end{figure}

Given a simple undirected graph $G=(V,E)$, the longest induced path problem (LIPP), also known as the maximum induced path problem, consists of obtaining a maximum cardinality subset of vertices $W\subseteq V$ inducing a simple path. 
More formally, denote by $G[W]=(W,E')$ the graph induced in $G$ by $W\subseteq V$, whose set of edges $E'\subseteq E$ is formed by all the edges in $E$ whose extremities belong to $W$, namely, $E'= \{e = uv \in E \ | \ u,v \in W\}$. 
LIPP consists of obtaining a maximum cardinality subset $W\subseteq V$ inducing a simple path $G[W]$. The problem is known to be NP-hard as its decision version is NP-complete~\cite{GarJoh79}. Figure~\ref{fig:longestinducedpath} exemplifies an input graph and one of its longest induced paths.

LIPP encounters applications in both practical and more graph theoretical situations.
Obtaining the longest induced paths in hypercube graphs is known as the snake-in-the-box problem and has applications in error-checking codes, communications, and data storage~\cite{Kau58,YehSch12,HooRecSawWon15}.
Given a graph and two predefined vertices $u,v \in V$, the detour distance between $u$ and $v$ is defined as the length of the longest induced path between them~\cite{ChaJohTia93}.
In this context, LIPP finds applicability in the evaluation of worst case transmission times in large communication and neural networks~\cite{Gav02}.
Additionally, it is also applicable to the analysis of social networks as it extends the concept of diameter of a graph, which is given by the longest of its shortest paths~\cite{MatVerProPas19}.
Furthermore, the existence of long induced paths plays an important role in the characterization of properties for several problems in graph theory~\cite{LozRau03,GolPauSon14,BonChuMacSchSteZho18,ChuSchSpiSteZho19,JafKwoTel20}.

Although LIPP is NP-hard in general, there are several classes of graphs for which it can be solved in polynomial time~\cite{Gav02,KraMulTod03,IshOtaYam08,JafKwoTel20}.
However, to the best of our knowledge,
approaches for solving general instances of the problem were only proposed very recently.
\citeA{MatVerProPas19} presented three compact integer programming (IP) formulations and an exact iterative IP-based algorithm using their IP 
formulations. The authors also presented a randomized heuristic to tackle larger instances of the problem.
\citeA{BokChiWagWie20,BokChiWagWie20b} described branch-and-cut approaches based on IP formulations using cutset (or generalized subtour elimination) constraints and clique inequalities.
The authors showed that the proposed formulations provide stronger linear relaxation bounds than those presented in \citeA{MatVerProPas19}. Furthermore, computational experiments demonstrated the superiority of the new formulations in terms of the number of instances solved to optimality and the median times for solving them. 
\citeA{MarRib21} proposed an exact backtracking algorithm, based on which they also derived a heuristic approach.

In addition, several problems which are somehow related to the longest induced path problem have been tackled using integer programming approaches in the literature.
\citeA{LjuWeiPfeKlaMutFis06} and \citeA{CosCorLap09} presented integer programming formulations and branch-and-cut methods for Steiner tree problems.
The problem of obtaining the longest induced simple cycle of a graph was considered in \citeA{LucSalSim}.
Formulations for the maximum weighted connected subgraph problem were proposed in~\citeA{AlvLjuMut13}, and \citeA{RehKoc19}.
\citeA{AgrDahHauPin17} analyzed exact approaches for finding maximum $k$-regular induced subgraphs of a graph. \citeA{MelQueRib21} considered the maximum weighted induced forest problem in the context of solving the minimum weighted feedback vertex set problem.
\rafaelC{\citeA{MelRib21}} proposed new IP formulations for the maximum weighted induced forest problem and showed how to adapt their approaches to find maximum weighted induced trees.

In the same line of the recent works of \citeA{MatVerProPas19}, \citeA{BokChiWagWie20,BokChiWagWie20b}, and \citeA{MarRib21}, we consider the longest induced path problem for general graphs.
We propose two new formulations with exponential numbers of constraints. 
We perform a theoretical comparison regarding the polyhedra defined by these formulations and that of a state-of-the-art formulation available in the literature~\cite{BokChiWagWie20}.
\rafaelC{
We also propose an effective branch-and-cut approach for the problem which, based on the characteristics of the instance, either adds \textit{a priori} to the formulation the clique inequalities for all the maximal cliques in the graph, or applies a cutting-plane heuristic to separate such inequalities.}
Extensive computational experiments are conducted, highlighting the superiority of our approaches when compared to a state-of-the-art formulation in terms of both the number of instances solved to optimality and the median times to solve them.

The remainder of the paper is organized as follows. 
Section~\ref{sec:formulations} describes the newly proposed formulations.
Section~\ref{sec:experiments} presents the computational experiments, where implementation details of the branch-and-cut approach are also given. Concluding
remarks are discussed in Section~\ref{sec:concluding}.
For the sake of completeness, the state-of-the-art formulation proposed in \citeA{BokChiWagWie20} is presented in Appendix~\ref{sec:germanformulation}.
A theoretical comparison of the polyhedra defined by the different, existing and new, problem formulations is presented in Appendix~\ref{sec:theoreticalcomparisonformulations}.
A theoretical analysis of the different clique inequalities is presented in Appendix~\ref{sec:comparecliqueinequalities}.

\section{New integer programming formulations}
\label{sec:formulations}

In this section, we propose two new formulations with an exponential number of constraints for the longest induced path problem (LIPP). We also describe the clique inequalities available for problems related to encountering induced subgraphs. Both formulations are undirected and consider a slightly modified graph constructed as follows, as \citeA{BokChiWagWie20} also did.
Given the graph $G=(V,E)$, we build $G_s=(V_s,E_s)$ with $V_s = V \cup \{s\}$ and $E_s = E \cup \{sv \ : \ v \in V \}$. The goal of the dummy vertex $s$ is to be linked to both extremities of the induced path $G[W]$, with $W\subseteq V$. 
In the remainder of the paper, let $E(V') \subseteq E$ be the set of edges in $E$ with both extremities in $V'$, and $\delta_G(V')$ be the set of edges in $G$ with an extremity in $V'$ and another one in $\bar{V'}= V\setminus V'$.

\subsection{Formulation with explicit cycle elimination constraints}
\label{sec:treecycleelimination}

In order to formulate LIPP as an integer program, define the binary variable $y_v$ to be equal to one if the vertex $v\in V_s$ belongs to the solution, 
zero otherwise. Besides, consider the binary variable $x_e$ to be equal to one if edge $e\in E_s$ is in the solution, 
zero otherwise.
Let $\mathcal{C}$ denote the family of all cycles in the graph $G$.
The formulation with explicit cycle elimination constraints can be defined as
\begin{align}
(\textrm{cec})  \qquad & \max \sum_{v \in V} y_{v}  \label{cyc:obj}\\
\quad &  \sum_{e \in \delta_{G_s}(v) } x_{e} = 2 y_v , \ \ \ & \forall \ v \in V,    \label{cyc:01} \\
& \sum_{e \in \delta_{G_s}(s) } x_{e} = 2,   \label{cyc:02} \\
& \sum_{v \in C} y_v \leq |C|-1, \ \ \ & \forall \ C \in \mathcal{C}, \label{cyc:03}\\
& x_{e} \leq y_v, \ \ \  & \forall \ v \in V,\ e \in \delta_{G_s}(v),  \label{cyc:04}\\
& x_{e} \geq  y_u + y_v - 1, \ \ \  & \forall \ e=uv \in E, \label{cyc:05}\\
& x \in \{0,1\}^{|E_s|}, \label{cyc:06}\\
&  y \in \{0,1\}^{|V_s|}. \label{cyc:07}
\end{align}
The objective function~\eqref{cyc:obj} maximizes the number of vertices in the induced path.
Constraints \eqref{cyc:01} guarantee that each selected vertex has degree two.
Constraint \eqref{cyc:02} ensures exactly two edges are adjacent to the dummy vertex.
Constraints \eqref{cyc:03} force the induced subgraph to be acyclic. Note that there is an exponential number of such constraints, one for each cycle in the graph. 
Constraints \eqref{cyc:04} and \eqref{cyc:05} ensure the path is induced.
Constraints \eqref{cyc:06} and \eqref{cyc:07} determine, respectively, the integrality of the $x$ and $y$ variables.

Note that, similarly to \rafaelC{\citeA{BokChiWagWie20}}, the formulation assumes $|E|>1$. We remark, though, that such assumption is not restrictive, as obtaining the optimal solution for a graph without any edges is straightforward.
An alternative way that could include the trivial graph as a feasible input would be to insert an additional dummy vertex to the transformed graph and, instead of closing a cycle with both extremities of the induced path, one would build a path between the two dummy vertices.

\subsection{Formulation with cutset constraints}
\label{sec:cutsetformulation}

The undirected cutset formulation is similar to that with explicit cycle elimination constraints, with the difference that it guarantees the elimination of cycles using cutset constraints. Using the same variables defined in Subsection~\ref{sec:treecycleelimination}, it can be cast as
\begin{align}
(\textrm{cut}) \qquad & \eqref{cyc:obj}-\eqref{cyc:02}, \eqref{cyc:04}-\eqref{cyc:07}, \nonumber\\ 
& \sum_{e \in \delta_{G_s}(S)} x_e \geq 2 y_v, \ \ \ & S \subseteq V, \ v \in {S}. \label{cutset:01}
\end{align}
Constraints~\eqref{cutset:01} guarantee the solution to be acyclic in $G$ by ensuring connectivity. To be more specific, given a partition $\{S,\bar{S}\}$, with vertex $v\in {S}$, the constraint enforces at least two edges in the cut $(S,\bar{S})$ to be in the solution whenever $y_v=1$.

Note that an alternative well-known approach for ensuring connectivity, which is known to be equivalent to the use of cutset inequalities, is the employment of generalized subtour elimination constraints~\cite{GoeMyu93,BokChiWagWie20}. This means that an equivalent way to ensure \eqref{cutset:01} is via
\begin{equation}
 \sum_{e \in E(S)} x_e \leq \sum_{u \in S \setminus \{v\}} y_u, \ \ \ S \subseteq V, \ v \in S. \label{subtour:01}
\end{equation}
Constraints \eqref{subtour:01} guarantee that for a given subset $S$ of vertices, the number of edges connecting them is at most the number of selected vertices minus one (whenever there is at least one selected edge from $E(S)$).

\subsection{Clique inequalities}

A clique in a graph is a subset of its vertices which are all pairwise adjacent.
Consider $\mathcal{K}$ to be the family of all cliques in the graph $G$.
\citeA{BokChiWagWie20,BokChiWagWie20b} described the clique inequalities using the edge variables as
\begin{equation}\label{ineq:clique}
    \sum_{e\in E(K)} x_e \leq 1, \ \ \ \forall \ K \in \mathcal{K}.
\end{equation}
Inequalities \eqref{ineq:clique} enforce the number of edges connecting vertices in a clique to be at most one.

On the other hand, clique inequalities can also be modeled using the variables corresponding to the vertices~\cite{BruMafTru00,MelRib21}, resulting in
\begin{equation}\label{ineq:cliquey}
    \sum_{v\in K} y_v \leq 2, \ \ \ \forall \ K \in \mathcal{K}.
\end{equation}
Inequalities \eqref{ineq:cliquey} ensure that at most two vertices in a clique are selected.

\subsection{Theoretical analysis of the formulations and clique inequalities}

We provide a theoretical comparison regarding the polyhedra defined by  formulations cec, cut, and BCWWy~\cite{BokChiWagWie20,BokChiWagWie20b} in Appendix~\ref{sec:theoreticalcomparisonformulations}. Namely, we show that the polyhedra defined by cut and BCWWy are equivalent. We also prove that the polyhedra determined by cut and BCWWy are strictly contained in that defined by cec. For completeness, BCWWy is described in Appendix~\ref{sec:germanformulation}.

A theoretical analysis of the different clique inequalities, \eqref{ineq:clique} and \eqref{ineq:cliquey}, is presented in Appendix~\ref{sec:comparecliqueinequalities}. We demonstrate that \eqref{ineq:clique} and \eqref{ineq:cliquey} are not equivalent when applied to our formulation. In addition, we characterize conditions that must hold for each inequality to imply the other.

\section{Computational experiments}
\label{sec:experiments}

In this section, we report the computational experiments assessing the performance of the newly proposed formulations.
The experiments were carried out on a machine running under Ubuntu, with an Intel(R) Core(TM) i7-8700 Hexa-Core 3.20 GHz and 16 GB of RAM. The formulations were coded in Julia v1.4.2, using JuMP v0.18.6. Furthermore, the formulations were solved using Gurobi 9.0.2.


We considered in our experiments the instances proposed for the longest induced path problem in \citeA{MatVerProPas19} and \citeA{BokChiWagWie20}, also used by~\citeA{MarRib21}. These instances can be obtained from~\citeA{instances}, where additional details can be encountered. 
They are grouped into four sets, which are: RWC, MG, BAS, and BAL.
RWC is composed of 22 real-world networks corresponding to communication and social networks of companies, characters in books, as well as transportation, biological and technical networks.
MG corresponds to \textit{The Movie Galaxy} dataset and contains 773 graphs associated to social networks of movie characters \cite{KamSchAlbZasHid18}.
BAS and BAL were generated by \citeA{BokChiWagWie20} using the Barabási-Albert probabilistic model for scale-free networks~\cite{BarAlb99} in an attempt to recreate those used in \citeA{MatVerProPas19}.
BAS is composed of 120 graphs divided into four subsets with $(|V|,d) \in \{(20,3),(30,3),(40,3),(40,2)\}$, where $|E|=(|V|-d)\times d$, having 30 instances each.  
BAL is composed of 150 graphs with 100 vertices divided into five subsets with $d \in \{2,3,10,30,50\}$, each of them containing 30 instances.

\subsection{Tested approaches}
\label{sec:testedapproaches}

The following approaches were considered in our experiments:
\begin{itemize}
    \item[-] cec: the formulation with explicit cycle elimination constraints, described in Section~\ref{sec:treecycleelimination}; 
    \item[-] cut: the formulation with cutset constraints, described in Section~\ref{sec:cutsetformulation};
    \item[-] $\mathrm{C^{n,c}_{int}}$: best performing formulation described in \citeA{BokChiWagWie20} (see Appendix~\ref{sec:germanformulation});
    \item[-] $\mathrm{C^{n}_{int}}$: formulation described in \citeA{BokChiWagWie20} corresponding to $\mathrm{C^{n,c}_{int}}$ without the clique inequalities (see Appendix~\ref{sec:germanformulation}).
\end{itemize}
Note that we do not explicitly consider in our experiments the approaches of \citeA{MatVerProPas19}, as they were already shown not to be as effective in general as those proposed in \citeA{BokChiWagWie20}.

The reported values for $\mathrm{C^{n,c}_{int}}$ and $\mathrm{C^{n}_{int}}$ are those 
in \citeA{BokChiWagWie20,BokChiWagWie20b}. We observe that we implemented their formulations and executed our implementations in our own computational environment. However, although our implementation showed a similar performance for the small and medium instances in the original benchmark set, its results were outperformed by those reported by the authors for the larger instances.
Thus, Table~\ref{tab:cpu-comparison} compares the computational resources involved in the experiments based on the benchmarks in~\citeA{cpubenchmark} to evaluate the performance of the formulations.


\begin{table}[H]
\centering
\small
\caption{CPU performance comparison with data extracted from \protect\citeA{cpubenchmark}: Higher values represent better performance. The second and third columns correspond to the hardware used in this paper and in \protect\citeA{BokChiWagWie20}, respectively.}
\label{tab:cpu-comparison}
\begin{tabular}{lrr}
\hline
\multicolumn{1}{c}{Benchmarks} & \multicolumn{1}{c}{Intel Core i7-8700} & \multicolumn{1}{c}{Intel Xeon Gold 6134} \\ \hline
Clock speed (GHz) & 3.2 & 3.2 \\
Turbo speed (GHz) & Up to 4.6 & Up to 3.7 \\
CPU single thread rating & 2,681 & 2,251 \\
CPU mark rating & 13,090 & 16,513 \\ \hline
\end{tabular}
\end{table}

\subsection{Implementation details and parameter settings}

We report in this section some relevant implementation issues.

\paragraph{(A) Separation of cycle constraints:} the separation of the cycle constraints \eqref{cyc:03} for formulation cec is performed based on the approach described in~\citeA{MelRib21}. It receives as input a separation graph $G_{sep}=G[V_{sep}]$ induced by the vertices corresponding to the $y$ variables assuming a nonzero value in the solution (which can be either a fractional solution corresponding to the linear relaxation or an integer solution). More specifically, $V_{sep} = \{v \in V_s \ | \ \hat{y}_v > 0\}$, where $\hat{y}_v$ represents the value assumed by variable $y_v$ in the solution. 
The separation of violated inequalities for integer solutions is performed using the well-known 
depth-first search algorithm (DFS)~\cite{Cor09} in $G_{sep}$. More specifically, for every back edge traversed during the DFS, the corresponding cycle is stored. The separation procedure adds to the formulation all the cycles encountered during the execution of DFS.
On the other hand, the separation for fractional solutions is performed heuristically. It uses an alternative DFS with certain greedy components in $G_{sep}$ by considering the vertices to be visited in non-increasing order of their associated $\hat{y}$  values. Whenever a back edge is traversed during the search, the algorithm checks if such cycle violates constraints \eqref{cyc:03}. In case that happens, it is stored. At the end of the execution of this alternative DFS, the separation procedure adds to the formulation all the violated inequalities which were encountered during the procedure.

\paragraph{(B) Separation of cutset constraints:} the separation of the cutset constraints \eqref{cutset:01} for formulation cut also takes $G_{sep}=G[V_{sep}]$ as input. The separation of violated inequalities for integer solutions is performed with a small variation of the well-known
breadth-first search algorithm (BFS)~\cite{Cor09} in $G_{sep}$. The algorithm starts at the dummy vertex $s$ and defines the part $\bar{S}$ of the partition $\{S,\bar{S}\}$ of $V_{sep}$ as the vertices which could be reached from $s$. In what follows, for each vertex in $v\in {S}$ the algorithm stores the corresponding violated inequality. At the end of the execution of the algorithm, all the encountered violated inequalities are added to the formulation.
Separation for fractional solutions is performed exactly using maximum flows (minimum cuts), following the approach of~\citeA{MagWol95}. Namely, the approach builds a directed graph based on the solution $(\hat{y},\hat{x})$ in which the capacities of the arcs are defined by the values assumed by the $\hat{x}$ variables. Thus, a maximum flow problem is solved from $s$ to each $v \in V_{sep}$. A violated constraint \eqref{cutset:01} is stored whenever it is encountered. At the end of the execution of the algorithm, all the obtained violated inequalities are inserted into the formulation.

\paragraph{(C) Separation of clique inequalities:} the separation of clique inequalities \eqref{ineq:cliquey} for formulations cec and cut also takes the separation graph $G_{sep}=G[V_{sep}]$ as input. It uses the heuristic approach described in~\citeA{MelRib21}, which works as follows.  
Firstly, the vertices in $V_{sep}$ are sorted in non-increasing order of their corresponding $\hat{y}$ values. In case of ties, they are considered in non-increasing order of their degree in $G_{sep}$. 
Next, a vertex adjacent to all others that 
were already chosen is greedily chosen to compose the clique under construction. 
These steps are repeated while a maximal clique in $G_{sep}$ was not yet obtained. 
Whenever a violated inequality is obtained for a maximal clique in $G_{sep}$, the approach attempts to lift such inequality by possibly adding new vertices that were not in the separation graph to get a maximal clique in the original graph $G$. 
The selection of vertices to lift the inequality uses a similar greedy idea, but it only considers their degrees.

\paragraph{(D) \textit{A priori} addition of clique inequalities:} in certain situations, we also consider adding \textit{a priori} to the formulation the clique inequalities corresponding to all the maximal cliques in the graph. The enumeration of maximal cliques can be performed based on the algorithm of Bron and Kerbosch~\cite{BroKer73,TomTanTak06}. Although the maximum number of such cliques can be exponential 
\cite{MooMos65}, it was observed in \citeA{BokChiWagWie20} that, for most of the benchmark instances, the number of maximal cliques was rather reasonably tractable. 

\paragraph{(E) Settings and further details:} whenever the number of maximal cliques with at least three vertices do not exceed a parameter $max_{cl}$, all the corresponding clique inequalities are added \textit{a priori} to the formulation. Otherwise, the separation of clique inequalities is employed.
Based on preliminary tests to check the tractability of larger formulations, in our experiments, $max_{cl}$ was set to 500. 
All the separation procedures were implemented as callbacks in the MIP solver. The separations for fractional solutions were configured to be executed only at the root node in an attempt not to overload the formulation with inequalities generated throughout the search tree. The MIP solver was executed using the standard configurations, except for the relative optimality tolerance gap, which was set to $10^{-6}$, and for the number of used threads, which was fixed to one. Each execution of the solver was limited to 1200 seconds. 

\subsection{Results}

In this section, we compare our formulations $\mathrm{cec}$ and $\mathrm{cut}$ with the state-of-the-art integer programming approaches proposed by~\citeA{BokChiWagWie20}, with one single thread for each run. 

Table~\ref{tab:RWC_running_times} displays the computational experiments on the RWC instances. 
The second column gives the optimal value. 
The third and fourth columns show the number of vertices and edges in each instance, respectively. 
The fifth and sixth columns display the time in seconds taken by the best $\mathrm{ILP_{Cut}}$ implementations ($\mathrm{C^n_{int}}, \mathrm{C^{n,c}_{int}}$) of \citeA{BokChiWagWie20}. 
The last two columns, indicated respectively by $\mathrm{cec}$ and $\mathrm{cut}$, present the running times of our formulations. 
The time limit was set to 1200 seconds and timeouts are denoted by \clock.

The table shows that, for the RWC instances, our new formulation $\mathrm{cec}$ outperformed the other implementations in terms of the number of instances solved to optimality. 
Formulation cec solved all but one instance, being the only one to solve the large instances \texttt{494bus} and \texttt{662bus}. 
The running time of cec is only 0.2\% and 10.6\% of those of formulations $\mathrm{C^n_{int}}$ and $\mathrm{C^{n,c}_{int}}$, respectively, for instance anna.
A more straightforward and fair comparison between the performances of the new formulations proposed in this work and those in~\citeA{BokChiWagWie20b} can be done considering the benchmarks in Table~\ref{tab:cpu-comparison}. The ratio $\frac{2681}{2251}\approx 1.19$ between the CPU single thread ratings of the machines used in each work gives a good approximation for the relative speed between them. If the running times for formulations cec and cut were adjusted by multiplying them by the ratio 1.19, under these conditions we could see that 
$\mathrm{cec}$ would be the fastest formulation for eleven RWC instances, followed by $\mathrm{cut}$ and $\mathrm{C^{n,c}_{int}}$ for seven and three ($\texttt{high-tech}$, $\texttt{karate}$ and $\texttt{usair}$) instances, respectively. 
Formulation $\mathrm{C^{n}_{int}}$ was never the fastest.

\begin{table}[H]
\centering
\small
\caption{Running times in seconds of the formulations on the RWC instances.}
\label{tab:RWC_running_times}
\begin{tabular}{lrrrrrrr}
\hline
\multicolumn{1}{l}{Instance} 
& \multicolumn{1}{c}{OPT} & \multicolumn{1}{c}{$|V|$} & \multicolumn{1}{c}{$|E|$} & \multicolumn{1}{c}{$\mathrm{C^{n}_{int}}$} 
& \multicolumn{1}{c}{$\mathrm{C^{n,c}_{int}}$}& \multicolumn{1}{c}{$\mathrm{cec}$} & \multicolumn{1}{c}{$\mathrm{cut}$} \\ \hline
high-tech & 13 & 33 & 91 & 0.51 & 0.41 & 0.68 & 0.61 \\
karate & 9 & 34 & 78 & 1.07 & 0.66 & 0.59 & 0.75 \\
mexican & 16 & 35 & 117 & 1.22 & 0.87 & 0.65 & 0.51 \\
sawmill & 18 & 36 & 62 & 0.85 & 0.82 & 0.55 & 0.45 \\
chesapeake & 16 & 39 & 170 & 2.29 & 3.19 & 0.72 & 0.56 \\
tailorS1 & 13 & 39 & 158 & 1.51 & 3.29 & 0.73 & 0.58 \\
tailorS2 & 15 & 39 & 223 & 3.20 & 2.89 & 0.96 & 0.81 \\
attiro & 31 & 59 & 128 & 1.20 & 0.89 & 0.64 & 0.47 \\
dolphins & 24 & 62 & 159 & 19.21 & 3.01 & 0.82 & 0.72 \\
krebs & 17 & 62 & 153 & 16.00 & 3.90 & 0.69 & 1.73 \\
prison & 36 & 67 & 142 & 3.62 & 1.02 & 0.57 & 0.44 \\
huck & 9 & 69 & 297 & 114.27 & 5.96 & 0.82 & 5.53 \\
sanjuansur & 38 & 75 & 144 & 8.22 & 3.79 & 0.72 & 0.81 \\
jean & 11 & 77 & 254 & 81.03 & 3.88 & 0.69 & 0.56 \\
david & 19 & 87 & 406 & 85.88 & 6.93 & 0.70 & 0.56 \\
ieeebus & 47 & 118 & 179 & 15.69 & 22.72 & 0.88 & \multicolumn{1}{c}{\clock} \\
sfi & 13 & 118 & 200 & 15.13 & 3.31 & 0.55 & 0.39 \\
anna & 20 & 138 & 493 & 439.23 & 7.09 & 0.75 & 0.92 \\
usair & 46 & 332 & 2126 & \multicolumn{1}{c}{\clock} & 922.94 & 887.76 & \multicolumn{1}{c}{\clock} \\
494bus & 142 & 494 & 586 & \multicolumn{1}{c}{\clock} & \multicolumn{1}{c}{\clock} & 88.41 & \multicolumn{1}{c}{\clock} \\
662bus & 305 & 662 & 906 & \multicolumn{1}{c}{\clock} & \multicolumn{1}{c}{\clock} & 212.54 & \multicolumn{1}{c}{\clock} \\
yeast & unknown & 2361 & 6646 & \multicolumn{1}{c}{\clock} & \multicolumn{1}{c}{\clock} & \multicolumn{1}{c}{\clock} & \multicolumn{1}{c}{\clock} \\ \hline
\# of timeouts &  &  &  & \multicolumn{1}{c}{4} & \multicolumn{1}{c}{3} & \multicolumn{1}{c}{1} & \multicolumn{1}{c}{5} \\ \hline
\end{tabular}%
\end{table}

\begin{figure}[H]
	\centering
		\includegraphics[scale=.85]{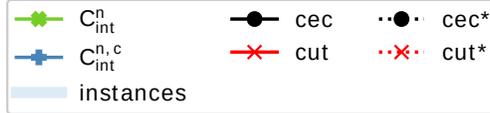}
	\caption{Identifications of the formulations.}
	\label{fig:legend}
\end{figure}

\begin{figure}[H]
	\centering
		\includegraphics[scale=.8]{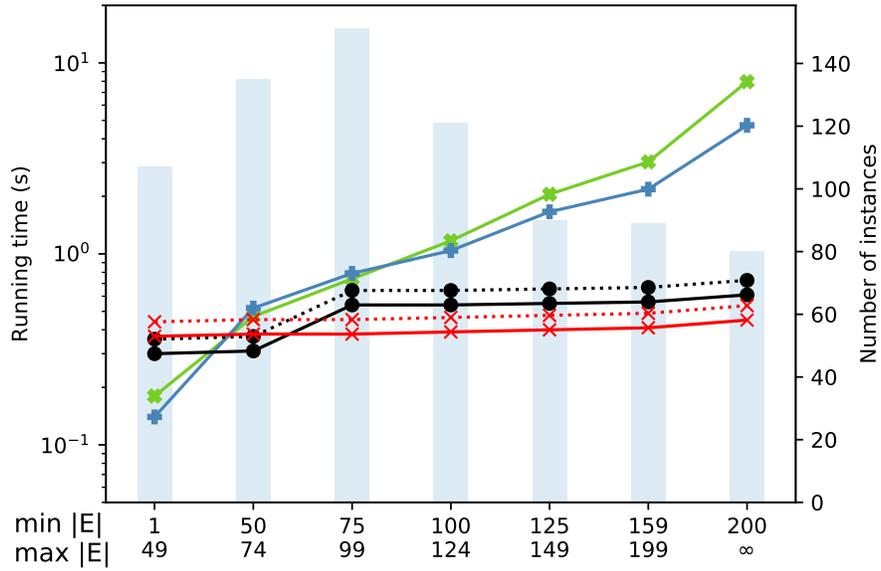} \\
(a) Running times in seconds.\\
		\includegraphics[scale=.8]{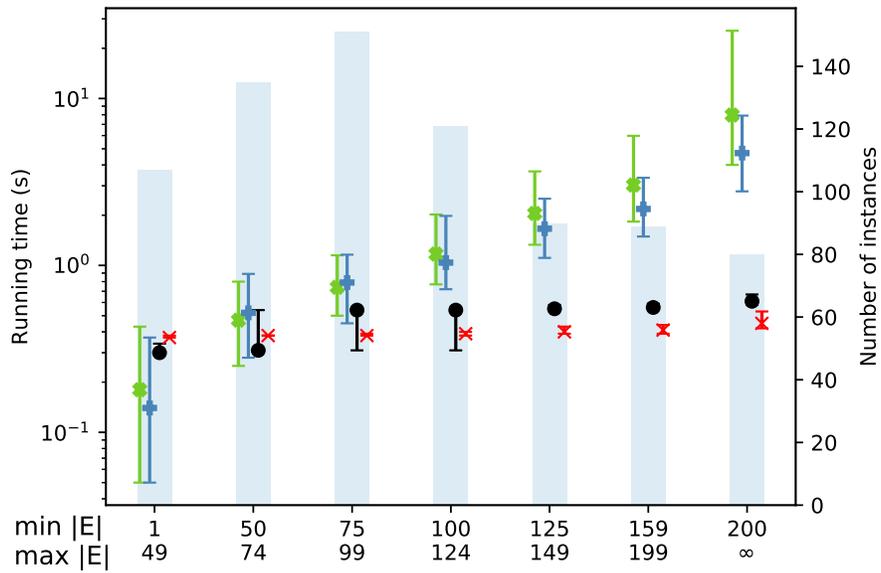}\\
(b) Boxplots of the running times in seconds.
\caption{Computational results on the MG instances. The new formulations cec and cut are more robust and their running times are much less dependent on the instance size.}
	\label{fig:results_MG_boxplot}
\end{figure}

\begin{figure}[H]
	\centering
		\includegraphics[scale=.8]{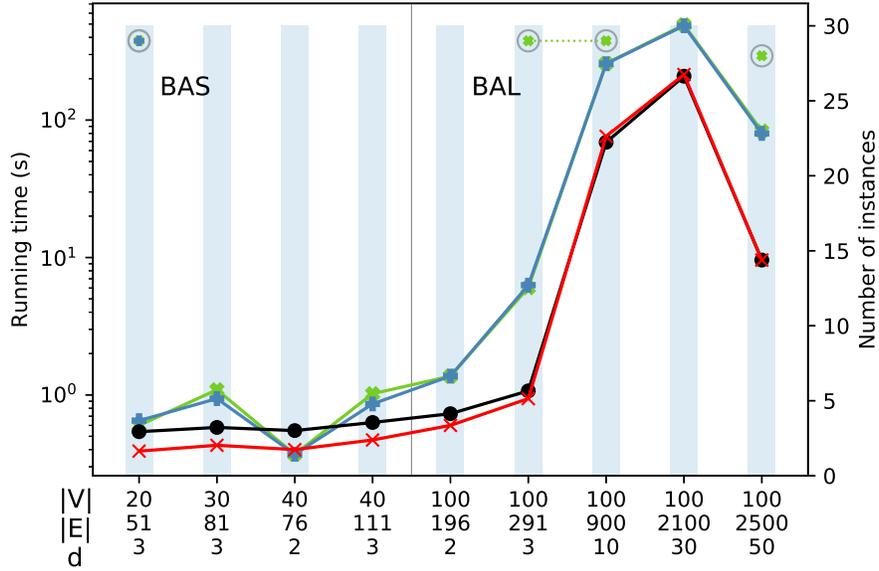} \\
		(a) Running times in seconds. \\
		\includegraphics[scale=.8]{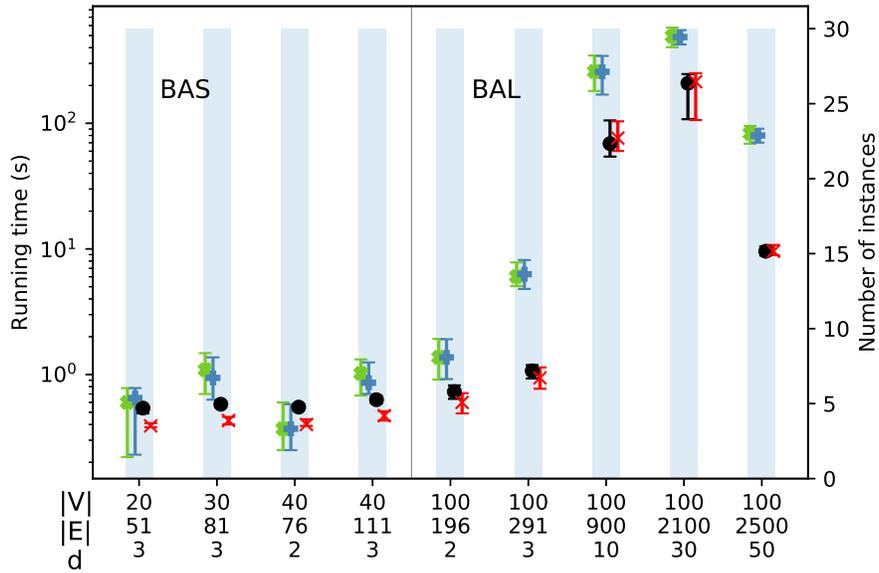}\\
		(b) Boxplots of the running times in seconds.
\caption{Computational results on the BAS and BAL  instances. Although the four formulations presented a similar behavior, the new formulations cec and cut showed smaller median times, in particular for the BAL instances.}	\label{fig:results_BAS-BAL_boxplot}
\end{figure}

\begin{figure}[H]
	\centering
		\includegraphics[scale=.8]{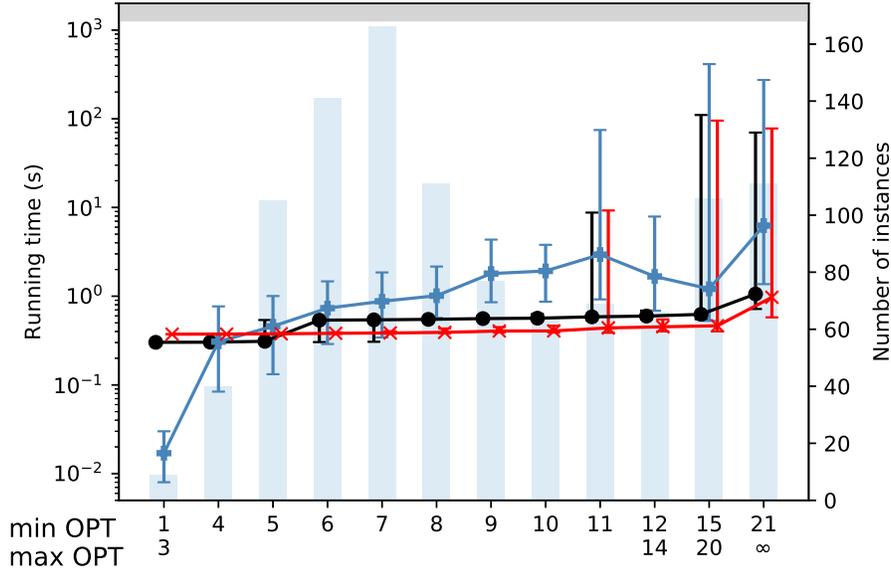}
	\caption{Running times vs. optimal values for all instances: whiskers mark the 20\% and 80\% percentiles. The gray area on top of the plot marks timeouts. Each run was limited to 20 minutes. The new formulations cec and cut presented much smaller variations in the running times and lower median times in most cases.}
	\label{fig:results_ALL_plot}
\end{figure}

Figure~\ref{fig:legend} shows the legend used in {Figures~\ref{fig:results_MG_boxplot}-\ref{fig:results_ALL_plot}, with the identification of our formulations (cec and cut) and of
the two best $\mathrm{ILP_{Cut}}$ implementations of \citeA{BokChiWagWie20}. 
We also indicate
by $\mathrm{cec^*}$ and $\mathrm{cut^*}$ the results obtained by formulations $\mathrm{cec}$ and $\mathrm{cut}$, respectively, with their running times multiplied by the factor $\frac{2681}{2251} \approx 1.19$.}

Figure~\ref{fig:results_MG_boxplot} displays comparative results for the MG instances, with the horizontal axis indicating the subsets into which the MG graphs were divided, according to their number of edges.
Figure~\ref{fig:results_BAS-BAL_boxplot} shows the comparative results for the instances in sets BAS and BAL, with the horizontal axis indicating the subsets into which the graphs were divided, according to their number of vertices, their number of edges and the value of parameter $d$. 
Figure~\ref{fig:results_ALL_plot} correlates the median running times of three formulations with the size of the longest induced path (OPT), considering all 1065 test instances.
The horizontal axis of the figure indicates the subsets into which the graphs were divided, according to the optimal value.
Vertical bars in light blue in the background give the number of instances in each subset.
For each formulation, we represent the median of the running times over all instances in the same subset. 
Furthermore, in Figures~\ref{fig:results_MG_boxplot}(b),~\ref{fig:results_BAS-BAL_boxplot}(b) and~\ref{fig:results_ALL_plot} the whiskers mark the 20\% and 80\% percentiles of the running times for each subset.
In the cases where
not all instances in the same subset have been solved to optimality, 
we indicate the number of solved instances by gray encircled markers connected by dotted lines (see Figure~\ref{fig:results_BAS-BAL_boxplot}(a)).

Figures~\ref{fig:results_MG_boxplot}-\ref{fig:results_ALL_plot} show that even though cec and cut can be outperformed by $\mathrm{C^{n}_{int}}$ and $\mathrm{C^{n,c}_{int}}$ for some of the small instances, they become much more effective than the latter as the instance sizes become larger. It is noticeable that in most cases our formulations present smaller variations in the running times.
We can see from Figure~\ref{fig:results_MG_boxplot} that for instances MG our formulations are more robust and their running times much less dependent on the instance sizes.
Figure~\ref{fig:results_BAS-BAL_boxplot} highlights the fact that, for instances BAS and BAL, the four formulations present a similar behavior regarding how the running times increase as the instances become larger, but it is noteworthy that our formulations cec and cut present much lower median times.


Table~\ref{tab:number_timeouts} summarizes, for each group of instances, the number of timeouts for each formulation considering the time limit of 1200 seconds.
Formulation cec, although less strong from a theoretical point of view, showed the best performance, being able to solve all the 1065 instances, except the largest one (the instance named $\texttt{yeast}$), which was not solved by any of the formulations.
\rafaelC{
A possible explanation for the good performance of cec is that it can still achieve good bounds (which are close to those obtained by cut) at the root node using lower computational effort for the considered instances.
Furthermore, we observed that the solver can frequently achieve good-quality feasible solutions earlier in the enumeration tree when using cec.
}

\begin{table}[H]
\centering
\small
\caption{Number of instances not solved to optimality by each formulation within the time limit of 1200 seconds.}
\label{tab:number_timeouts}
\begin{tabular}{lrrrrr}
\hline
\multicolumn{2}{c}{Group} & \multicolumn{4}{c}{Number of timeouts} \\ \hline
\multicolumn{1}{c}{Name} & \multicolumn{1}{c}{Instances} & \multicolumn{1}{c}{$\mathrm{C^{n}_{int}}$} & \multicolumn{1}{c}{$\mathrm{C^{n,c}_{int}}$} & \multicolumn{1}{c}{cec} & \multicolumn{1}{c}{cut} \\ \hline
RWC & 22 & 4 & 3 & \textbf{1} & 5 \\
MG & 773 & \textbf{0} & \textbf{0} & \textbf{0} & \textbf{0} \\
BAS & 120 & 1 & 1 & \textbf{0} & \textbf{0} \\
BAL & 150 & 4 & \textbf{0} & \textbf{0} & \textbf{0} \\ \hline
Total & 1065 & 9 & 4 & \textbf{1} & 5 \\ \hline
\end{tabular}
\end{table}

\subsection{Experiments with more challenging larger instances}
\label{subsec:experiments_warmstarts}

In this section we analyze the performance of the newly proposed approaches, cec and cut, on a benchmark set composed of more challenging, larger instances.
The goal of these additional experiments was twofold. Firstly, we wanted to show that even though our approaches were able to solve nearly all the original benchmark instances to optimality, there are still instances that are hard to solve in practice. Secondly, we investigated the impact of offering warm start solutions (i.e., initial feasible solutions) to the formulations. 

This benchmark set is composed of 23 challenging larger instances.
It contains: 
(a) three hypercube graphs, where $k$-cube denotes the $k$-dimensional hypercube with $2^k$ vertices; 
(b) six 2-connected random graphs, originally proposed by~\citeA{CarCerGenPar2011} for the minimum weighted feedback vertex set problem (MWFVS), each of which identified in our work as Rand\_$|V|$\_$|E|$\_seed; 
(c) three toroidal graphs, also proposed for MWFVS, each of them represented by Toro\_$n$\_$n$\_seed, where $n$ represents the dimension of a square grid based on which the toroidal graph was obtained (notice that each toroidal graph has $n\times n$ vertices);
(d) ten larger BAL instances, half of them with 1000 vertices, 9900 edges, and $d = 10$, while the other five have 1485 vertices, 13860 edges and $d = 3$;
and (e) the largest RWC instance \textit{yeast}, which was not solved by any of the formulations within 1200 seconds.
Note that we consider the unweighted versions of the instances corresponding to items (b) and (c) enumerated in this paragraph.

The warm starts used in the experiments were produced by the \mbox{G-HLIPP} heuristic of~\citeA{MarRib21}.
This greedy heuristic explores all vertices of the graph as possible source vertices of induced paths.
The source vertices are selected in the non-increasing order of their eccentricities and ties are broken in favor of vertices with smaller degrees. 
Parameter $maxpaths$ of the heuristic, which limits the number of induced paths that are explored from each source vertex, was empirically set to 5000.
The heuristic stops after generating a sequence of $maxpaths$ induced paths from each vertex of the graph that do not improve the incumbent solution.

Given the large sizes of the instances, the maximum allowed running time for the experiments in this section was set to 3600 seconds (1 hour) for each run.
For the executions of the plain formulation, i.e., without warm starts, the solver was executed with the full time limit of 3600 seconds.
For the executions with warm starts, the heuristic \mbox{G-HLIPP} was run for 360 seconds (6 minutes, corresponding to 10\% of the total maximum allowed time), while the remaining 3240 seconds (54 minutes) were made available to run the formulation using the solver.

Table~\ref{tab:cec_cut_warmstart_timelimit3600} reports the computational results for formulations cec, cut, and their variants with warm starts.
The table shows that, among all formulations, cec with warm starts reached the lowest average gap and the highest number of best values regarding the incumbent (column Obj) and relative gap, but cec was better in terms of the number of best upper bounds (although by just one instance of difference). 
For eight instances none of the formulations with warm starts was able to improve in 54 minutes the best solution found by heuristic G-HLIPP and for eight instances the heuristic found better solutions than cec and cut.
We remark that the optimal solutions for instances 7-cube (51) and 8-cube (99) are known and were obtained in the context of specialized approaches for the snake-in-the-box problem~\cite{OstPet15}.

When we make a specific comparison between cec and cec with warm starts, the table shows that cec with warm starts reached the highest number of best values regarding the incumbent (column Obj) and the relative gap, but cec was better in terms of the number of best upper bounds (although by just an instance of difference). 
For eight instances cec with warm starts was not able to improve in 54 minutes the best solution found by heuristic G-HLIPP and for eight instances the heuristic found better solutions than cec.
When comparing cut and cut with warm starts, it can be noticed that cut with warm starts reached the highest number of best values regarding the incumbent (column Obj), upper bound, and relative gap. 
For ten instances cut with warm starts was not able to improve in 54 minutes the best solution found by heuristic G-HLIPP and for twelve instances the heuristic found better solutions than cut.

\begin{figure}[H]
	\centering
	\includegraphics[scale=.8]{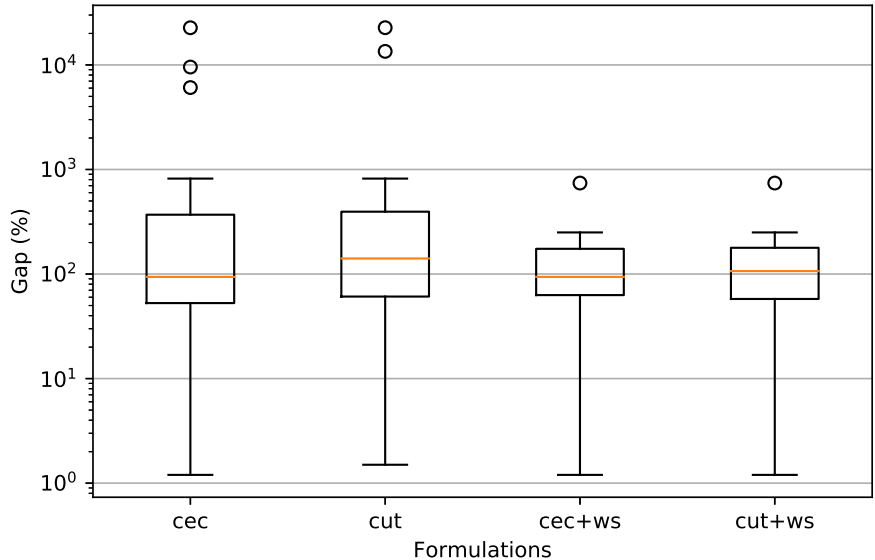}
    \caption{Boxplots of the relative gaps (\%) associated with our formulations cec, cut, and their variants with warm starts on the 23 difficult instances.
The box extends from the lower (25th percentile) to upper quartile (75th percentile) values of the data, with a line at the median. 
The whiskers extend from the box to show the range of the data. 
Beyond the whiskers, data were considered outliers (points outside 1.5 times the interquartile range) and were plotted as individual points.
We can see that the variants with warm starts presented less outliers and lower variability of the relative gaps and the interquartile range, confirming their greater robustness in comparison with the plain formulations cec and cut.
    }
	\label{fig:formulations_gap_boxplot}
\end{figure}

To summarize, the benchmark set considered in this section is composed of very challenging instances. The results indicate that even though nearly all original benchmark instances are solved, there is still room for advances in these approaches to solve these larger instances.
Furthermore, the use of warm starts was shown to be a very important contribution when compared to the use of the plain formulations, as they became much more stable when it comes to the achieved optimality gaps. This is also depicted in Figure~\ref{fig:formulations_gap_boxplot}. It can be noticed that the improvements were more significant for formulation cut. 


\newgeometry{left=1cm,right=1cm,bottom=2cm,top=2cm}
\begin{landscape}
\begin{table}[H]
\centering
\small
\caption{Performance of formulations cec, cut, and their variants with warm starts on difficult instances. The time limit for cec (resp. cut), G-HLIPP, and cec with warm starts (resp. cut with warm starts) was set to 3600, 360, and 3240 seconds, respectively. The best values are highlighted in bold.}
\label{tab:cec_cut_warmstart_timelimit3600}
\resizebox{1.2\textwidth}{!}{%
\begin{tabular}{lrrrrrrrrrrrrrrr}
\hline
\multicolumn{1}{c}{Instance} & \multicolumn{1}{c}{$|V|$} & \multicolumn{1}{c}{$|E|$} & \multicolumn{3}{c}{cec} & \multicolumn{3}{c}{cut} & \multicolumn{1}{c}{G-HLIPP} & \multicolumn{3}{c}{cec + warm start} & \multicolumn{3}{c}{cut + warm start} \\ \cline{4-9} \cline{11-16} 
\multicolumn{1}{c}{} & \multicolumn{1}{c}{} & \multicolumn{1}{c}{} & \multicolumn{1}{c}{Obj} & \multicolumn{1}{c}{UB} & \multicolumn{1}{c}{Gap (\%)} & \multicolumn{1}{c}{Obj} & \multicolumn{1}{c}{UB} & \multicolumn{1}{c}{Gap (\%)} & \multicolumn{1}{c}{Obj} & \multicolumn{1}{c}{Obj} & \multicolumn{1}{c}{UB} & \multicolumn{1}{c}{Gap (\%)} & \multicolumn{1}{c}{Obj} & \multicolumn{1}{c}{UB} & \multicolumn{1}{c}{Gap (\%)} \\ \hline
{7-cube} & 128 & 448 & \textbf{50} & 67 & \textbf{34.0} & 48 & 67 & 39.6 & 48 & 49 & \textbf{66} & 34.7 & 49 & 67 & 36.7 \\
{8-cube} & 256 & 1,024 & \textbf{88} & \textbf{139} & \textbf{58.0} & 86 & \textbf{139} & 61.6 & 84 & 86 & 140 & 62.8 & 87 & 140 & 60.9 \\
{9-cube} & 512 & 2,304 & 143 & 285 & 99.3 & 145 & 285 & 96.6 & \textbf{160} & \textbf{160} & \textbf{284} & \textbf{77.5} & \textbf{160} & \textbf{284} & \textbf{77.5} \\
{Rand\_200\_796\_10203} & 200 & 796 & \textbf{86} & \textbf{101} & \textbf{17.4} & 84 & \textbf{101} & 20.2 & 67 & \textbf{86} & 102 & 18.6 & 85 & \textbf{101} & 18.8 \\
{Rand\_200\_3184\_11203} & 200 & 3,184 & 32 & \textbf{62} & 93.8 & \textbf{33} & \textbf{62} & \textbf{87.9} & 30 & 30 & \textbf{62} & 106.7 & 30 & \textbf{62} & 106.7 \\
{Rand\_200\_12139\_12283} & 200 & 12,139 & 11 & 56 & 409.1 & 11 & 56 & 409.1 & \textbf{12} & \textbf{12} & \textbf{42} & \textbf{250.0} & \textbf{12} & \textbf{42} & \textbf{250.0} \\
{Rand\_300\_1644\_12739} & 300 & 1,644 & \textbf{101} & \textbf{149} & \textbf{47.5} & 93 & \textbf{149} & 60.2 & 82 & 92 & 150 & 63.0 & 97 & 150 & 54.6 \\
{Rand\_300\_7026\_13787} & 300 & 7,026 & 35 & 92 & \textbf{162.9} & \textbf{36} & 99 & 175.0 & 33 & 34 & \textbf{91} & 167.6 & 33 & \textbf{91} & 175.8 \\
{Rand\_300\_27209\_14811} & 300 & 27,209 & 11 & \textbf{101} & 818.2 & 11 & \textbf{101} & 818.2 & \textbf{12} & \textbf{12} & \textbf{101} & \textbf{741.7} & \textbf{12} & \textbf{101} & \textbf{741.7} \\
{Toro\_10\_10\_1187} & 100 & 200 & \textbf{59} & \textbf{61} & \textbf{3.4} & \textbf{59} & 62 & 5.1 & 54 & \textbf{59} & \textbf{61} & \textbf{3.4} & \textbf{59} & 62 & 5.1 \\
{Toro\_20\_20\_2283} & 400 & 800 & \textbf{259} & \textbf{262} & \textbf{1.2} & \textbf{259} & 263 & 1.5 & 213 & \textbf{259} & \textbf{262} & \textbf{1.2} & \textbf{259} & \textbf{262} & \textbf{1.2} \\
{Toro\_23\_23\_2595} & 529 & 1,058 & \textbf{344} & \textbf{349} & \textbf{1.5} & \textbf{344} & \textbf{349} & \textbf{1.5} & 282 & \textbf{344} & \textbf{349} & \textbf{1.5} & \textbf{344} & \textbf{349} & \textbf{1.5} \\
ba.1000.10.0 & 1,000 & 9,900 & 232 & \textbf{511} & 120.3 & 213 & 513 & 140.8 & 130 & \textbf{269} & \textbf{511} & \textbf{90.0} & 236 & 512 & 116.9 \\
ba.1000.10.1 & 1,000 & 9,900 & 271 & 513 & 89.3 & \textbf{278} & 513 & \textbf{84.5} & 135 & 272 & \textbf{512} & 88.2 & 271 & 513 & 89.3 \\
ba.1000.10.2 & 1,000 & 9,900 & 263 & \textbf{510} & 93.9 & 241 & 512 & 112.4 & 123 & 232 & \textbf{510} & 119.8 & \textbf{268} & 511 & \textbf{90.7} \\
ba.1000.10.3 & 1,000 & 9,900 & \textbf{304} & \textbf{511} & \textbf{68.1} & 208 & 515 & 147.6 & 128 & 264 & 512 & 93.9 & 284 & 514 & 81.0 \\
ba.1000.10.4 & 1,000 & 9,900 & \textbf{270} & \textbf{511} & \textbf{89.3} & 107 & 515 & 381.3 & 131 & 169 & \textbf{511} & 202.4 & 205 & 513 & 150.2 \\
ba.1000.3.0 & 1,485 & 13,860 & 7 & 675 & 9542.9 & 168 & \textbf{674} & 301.2 & \textbf{252} & \textbf{252} & 675 & \textbf{167.9} & \textbf{252} & 675 & \textbf{167.9} \\
ba.1000.3.1 & 1,485 & 13,860 & 11 & \textbf{679} & 6072.7 & 156 & 680 & 335.9 & 253 & \textbf{254} & 680 & \textbf{167.7} & \textbf{254} & 680 & \textbf{167.7} \\
ba.1000.3.2 & 1,485 & 13,860 & 158 & \textbf{679} & 329.7 & 134 & 681 & 408.2 & \textbf{234} & \textbf{234} & 680 & \textbf{190.6} & \textbf{234} & 680 & \textbf{190.6} \\
ba.1000.3.3 & 1,485 & 13,860 & 3 & \textbf{683} & 22666.7 & 3 & 684 & 22700.0 & \textbf{241} & \textbf{241} & \textbf{683} & \textbf{183.4} & \textbf{241} & \textbf{683} & \textbf{183.4} \\
ba.1000.3.4 & 1,485 & 13,860 & 105 & \textbf{680} & 547.6 & 84 & 684 & 714.3 & \textbf{243} & \textbf{243} & 682 & \textbf{180.7} & \textbf{243} & 682 & \textbf{180.7} \\
yeast & 2,361 & 6,646 & 226 & 516 & 128.3 & 4 & 543 & 13475.0 & 188 & \textbf{286} & \textbf{509} & \textbf{78.0} & 188 & 646 & 243.6 \\ \hline
Number of best values &  &  & 9 & \textbf{16} & 10 & 6 & 7 & 3 & 7 & \textbf{14} & 15 & \textbf{13} & 12 & 9 & 11 \\
Average gap (\%) &  &  &  &  & 1804.1 &  &  & 1764.3 &  &  &  & \textbf{134.4} &  &  & 138.8 \\ \hline
\end{tabular}%
}
\end{table}

\end{landscape}
\restoregeometry


\section{Concluding remarks}
\label{sec:concluding}

In this paper, we proposed two new formulations with an exponential number of constraints for the longest induced path problem on general graphs, together with effective branch-and-cut procedures for its solution. 
We proved that the polyhedra defined by formulation cut (wich ensures connectivity via cutset constraints) and by a state-of-the-art formulation, recently proposed in the literature~\cite{BokChiWagWie20}, are equivalent. Besides, we showed that they are strictly contained in the polyhedron defined by formulation cec (which is based on constraints that explicitly eliminate cycles). 
We also analyzed the strength of clique inequalities when using variables corresponding to edges and to vertices.

Extensive computational experiments showed that our newly proposed approach based on the explicit elimination of cycles, although less strong in theory, performs very well in practice. More specifically, it outperforms the others as it is able to solve all but one of the 1065 benchmark instances used so far in the literature, showing a clear advantage especially for the more challenging instances.

In addition, we performed extended computational experiments on a newly proposed benchmark set consisting of 23 hard-to-solve larger instances, which poses a tough challenge to our approaches. 
These extended experiments also showed that offering initial feasible solutions to our formulations had a very positive impact in reducing the variability of the achieved optimality gaps.

\section*{Acknowledgments}

Work of Rusl\'an G. Marzo was supported by a scholarship from Coordena\c{c}\~ao de Aperfeiçoamento de Pessoal de N\'\i vel Superior (CAPES) and by scholarship E-26/200.330/2020 from Fundação de Amparo à Pesquisa do Estado do Rio de Janeiro (FAPERJ).
Work of Rafael A. Melo was supported by the State of Bahia Research Foundation (FAPESB) and the Brazilian National Council for Scientific and Technological Development (CNPq). 
Work of Celso C. Ribeiro was partially supported by CNPq research grants
303958/2015-4, 425778/2016-9, and  and by FAPERJ research grant E-26/202.854/2017.
This work was also partially sponsored by CAPES, under Finance Code 001.

\bibliographystyle{apacite}
\bibliography{main}

\newpage

\begin{appendices}

\section{State-of-the-art integer programming formulation}
\label{sec:germanformulation}

The state-of-the-art integer programming formulation for the longest induced path problem was proposed by~\citeA{BokChiWagWie20,BokChiWagWie20b}. It considers the transformed graph $G_s=(V_s,E_s)$, with $V_s = V \cup \{s\}$ and $E_s = E \cup \{sv \ : \ v \in V \}$. Given the $x$ variables described in Section~\ref{sec:formulations}, their base formulation can be defined as
\begin{align}
(\textrm{BCWW})  \qquad & \max \sum_{e \in E} x_{e}  \label{ger:obj}\\
\quad & \sum_{e \in \delta_{G_s}(s) } x_{e} = 2,  &    \label{ger:02} \\
& \sum_{e \in \delta_{G_s}(v)} x_e \leq \sum_{e \in \delta_{G_s}(S)} x_e, \ \ \  & \forall \ S \subseteq V, \  v \in S, \label{ger:03}\\
& 2 x_{e} \leq \sum_{f \in \delta_{G_s}(\{u,v\})} x_f \leq 2, \ \ \  & \forall \ e=uv \in E,  \label{ger:04}\\
& x \in \{0,1\}^{|E_s|}. \label{ger:05}
\end{align}
The authors have observed that constraints~\eqref{ger:03} are equivalent to the generalized subtour elimination constraints \eqref{subtour:01}.
They also proposed a modified formulation using the $y$ variables (described in Section~\ref{sec:formulations}) and relaxes the integrality requirements on the $x$ variables. It can be cast as
\begin{align}
(\textrm{BCWWy}) \qquad & \eqref{ger:obj}-\eqref{ger:04}\\ 
& y_v = \frac{1}{2} \sum_{e \in \delta_{G_s}(v)} x_{e}, \ \ \  & \forall \ v \in V,  \label{gery:01}\\
& x \in [0,1]^{|E_s|}, \label{gery:02}\\
& y \in \{0,1\}^{|V|}. \label{gery:03}
\end{align}

We remark that~\citeA{BokChiWagWie20,BokChiWagWie20b} tested several configurations of their formulation. We refer to $\mathrm{C^{n,c}_{int}}$ as the best performing approach in their computational experiments, considering the number of instances solved to optimality and the average times for solving them. $\mathrm{C^{n,c}_{int}}$ uses formulation BCWWy, add the clique inequalities \eqref{ineq:clique} corresponding to all maximal cliques \textit{a priori} to the formulation, and separates the cutset inequalities \eqref{ger:03} only for integer solutions. Additionally, $\mathrm{C^{n}_{int}}$ denotes the variant of $\mathrm{C^{n,c}_{int}}$ without the addition of clique inequalities.

\newpage

\section{Theoretical comparison of the formulations cec, cut and BCWWy}
\label{sec:theoreticalcomparisonformulations}

Denote by $Q^{cec}$, $Q^{cut}$, and $Q^{BCWWy}$ the polyhedra defined by the linear relaxations of formulations cec, cut, and BCWWy, respectively. Each of these polyhedra is defined as the set of feasible points satisfying all the corresponding constraints. 
We remark that the objective functions \eqref{cyc:obj} and \eqref{ger:obj} are equivalent, even for the linear relaxations of the formulations.

\subsection{Comparing $Q^{cec}$ with $Q^{cut}$}

In what follows, we show that $Q^{cut} \subset Q^{cec}$. 
Consider a solution $(\hat{x},\hat{y}) \in Q^{cut}$. The proof consists in showing that $(\hat{x},\hat{y}) \in Q^{cec}$ and that there is a solution $(\bar{x},\bar{y}) \in Q^{cec}$ such that $(\bar{x},\bar{y}) \notin Q^{cut}$.
Firstly, notice that the only difference between these two formulations is related to the constraints to remove cycles, namely, \eqref{cyc:03} for cec and \eqref{cutset:01} for cut.

\begin{lemma}\label{lemma:cec-cut-1}
    Solution $(\hat{x},\hat{y})$ satisfies \eqref{cyc:03}.
\end{lemma}
\begin{proof}
    First, remember that constraints \eqref{cutset:01} imply \eqref{subtour:01}\cite{GoeMyu93,BokChiWagWie20}. Consider any cycle $C$. As \eqref{subtour:01} implies that $\sum\limits_{e \in E(C)} x_e \leq \sum\limits_{u \in C \setminus \{v\}} y_u$ for every $\rafaelC{v}\in C$, any right-hand side is at most $|C| - 1$. Thus, $(\hat{x},\hat{y})$ satisfies \eqref{cyc:03}. 
\end{proof}

\begin{lemma}\label{lemma:cec-cut-2}
    $Q^{cut} \subseteq Q^{cec}$.
\end{lemma}
\begin{proof}
The proof follows from Lemma~\ref{lemma:cec-cut-1} together with the fact that all other constraints defining these polyhedra are the same.
\end{proof}

\begin{figure}[ht!]
    \begin{center}
\subfigure[]{
	\begin{tikzpicture}[node distance=2.5cm,scale=1.0, every node/.style={scale=1.0}]
	\tikzstyle{vertex}=[circle, draw,line width=1pt]
	\tikzstyle{vertex2}=[circle, draw,line width=1pt,dashdotdotted]
	\tikzstyle{inv}=[circle]
	\tikzstyle{edge} = [draw,thick,-,line width=1pt]
	\tikzstyle{edge2} = [draw,thick,-,line width=1pt,dashdotdotted]
	
	\node[vertex] (c) {$c$};
	\node[vertex] [above left of = c] (a) {$a$};
	\node[vertex] [below left of = c] (b) {$b$};
    \node[vertex] [right of = c] (d) {$d$};
	\node[vertex] [below right of = d] (f) {$f$};
	\node[vertex] [above right of = d] (e) {$e$};
	
	\path[edge] (a) edge [above]node{} (b);        
	\path[edge] (a) edge [above]node{} (c);
	\path[edge] (b) edge [above]node{} (c);
	\path[edge] (c) edge [above]node{} (d);
	\path[edge] (d) edge [above]node{} (e);
	\path[edge] (d) edge [above]node{} (f);
	\path[edge] (e) edge [above]node{} (f);
	\end{tikzpicture}
} \hspace{2 cm}
\subfigure[]{
	\begin{tikzpicture}[node distance=2.5cm,scale=1.0, every node/.style={scale=1.0}]
	\tikzstyle{vertex}=[circle, draw,line width=1pt]
	\tikzstyle{vertex2}=[circle, draw,line width=1pt,dashdotdotted]
	\tikzstyle{inv}=[circle]
	\tikzstyle{edge} = [draw,thick,-,line width=1pt]
	\tikzstyle{edge2} = [draw,thick,-,line width=1pt,dashdotdotted]
	
	\node[vertex] (c) {$c$};
	\node[inv] [right of = c, node distance=0.6cm] (cl) {$\frac{2}{3}$};
	\node[vertex] [above left of = c] (a) {$a$};
	\node[inv] [above of = a, node distance=0.6cm] (al) {$\frac{2}{3}$};
	\node[vertex] [below left of = c] (b) {$b$};
	\node[inv] [below of = b, node distance=0.6cm] (bl) {$\frac{2}{3}$};
    \node[inv] [right of = c] (d) {};
	\node[vertex] [below right of = d] (f) {$f$};
	\node[inv] [below of = f, node distance=0.6cm] (dl) {$1$};
	\node[vertex] [above right of = d] (e) {$e$};
	\node[inv] [above of = e, node distance=0.6cm] (el) {$1$};
	\node[vertex2] [right of = d, node distance=4cm] (s) {$s$};
	
    \path[edge] (a) edge [left]node{$\frac{2}{3}$} (b);        
	\path[edge] (a) edge [above]node{$\frac{2}{3}$} (c);
	\path[edge] (b) edge [above]node{$\frac{2}{3}$} (c);
	\path[edge] (e) edge [left]node{$1$} (f);
	\path[edge2] (s) edge [above]node{$1$} (e);
	\path[edge2] (s) edge [above]node{$1$} (f);
	\end{tikzpicture}
}
\end{center}
   \caption{Examples of (a) an input graph $G$ with node set $V=\{a,b,c,d,e,f\}$ and (b) a fractional solution which also includes the dummy vertex $s$ depicting the nonzero variables $y_a = y_b = y_c = \frac{2}{3}$, $x_{ab}=x_{bc}=x_{ac}= \frac{2}{3}$, $y_e = y_f = 1$, and $x_{ef}=x_{es}=x_{fs}= 1$.}\label{fig:fractionalsolution}
\end{figure}
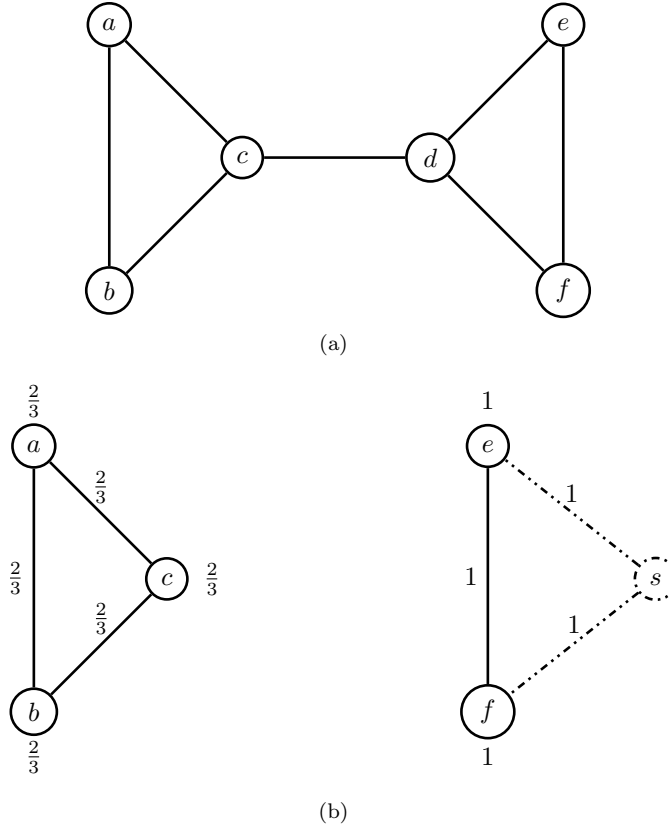

\begin{lemma}\label{lemma:cec-cut-3}
    $Q^{cec} \not\subseteq Q^{cut}$.
\end{lemma}
\begin{proof}
Consider the fractional solution $(\bar{x},\bar{y}) \in Q^{cec}$ represented by Figure~\ref{fig:fractionalsolution}. Note that it violates constraints \eqref{cutset:01} and, in consequence, $(\bar{x},\bar{y}) \notin Q^{cut}$.
\end{proof}

\begin{proposition}\label{proposition:cec-cut-4}
    $Q^{cut} \subset Q^{cec}$.
\end{proposition}
\begin{proof}
The proof follows from Lemmas~\ref{lemma:cec-cut-2}-\ref{lemma:cec-cut-3}.
\end{proof}

\subsection{Comparing $Q^{BCWWy}$ with $Q^{cut}$}

In the following, we demonstrate that $Q^{cut} = Q^{BCWWy}$. 
Consider a solution $(\hat{x},\hat{y}) \in Q^{cut}$. We show that $(\hat{x},\hat{y}) \in Q^{BCWWy}$. After that, we show that any solution $(\bar{x},\bar{y}) \in Q^{BCWWy}$ also belongs to $Q^{cut}$.

\begin{lemma}\label{lemma:bcwwy-cut-1}
    Solution $(\hat{x},\hat{y})$ satisfies \eqref{ger:02}, \eqref{gery:01}, \eqref{gery:02} and the continuous relaxation of \eqref{gery:03}.
\end{lemma}
\begin{proof}
    This is a straightforward implication of constraints \eqref{cyc:01}, \eqref{cyc:02}, and the continuous relaxations of \eqref{cyc:06} and \eqref{cyc:07}.
\end{proof}

\begin{lemma}\label{lemma:bcwwy-cut-2}
    Solution $(\hat{x},\hat{y})$ satisfies \eqref{ger:03}.
\end{lemma}
\begin{proof}
Constraints \eqref{cutset:01} imply that
\begin{equation*}
    \sum_{e \in \delta_{G_s}(S)} x_e \geq 2 y_v = \sum_{e \in \delta_{G_s}(v) } x_e. 
\end{equation*}
Therefore, $(\hat{x},\hat{y})$ satisfies \eqref{ger:03}.
\end{proof}

\begin{lemma}\label{lemma:bcwwy-cut-3}
    Solution $(\hat{x},\hat{y})$ satisfies \eqref{ger:04}.
\end{lemma}
\begin{proof}
Assume without loss of generality that $y_v \leq y_u$.  Summing up inequalities \eqref{cyc:01} for $u$ and $v$, we get
    \begin{equation}\label{eq:sumforuandv}
        \sum_{f \in \delta_{G_s}(\{u,v\})} x_f + 2 x_e = 2y_v + 2y_u.
    \end{equation}
    
    Firstly, we show that the inequality on the left is satisfied, namely, that $$2 x_{e} \leq \sum_{f \in \delta_{G_s}(\{u,v\})} x_f.$$ Observe that constraints
    \eqref{cyc:04} imply that $2x_e \leq 2 y_v$. This, together with \eqref{eq:sumforuandv}, ensures that
    \begin{equation}
    \sum_{f \in \delta_{G_s}(\{u,v\})} x_f \geq 2  y_u \geq 2x_e.
    \end{equation}
    
    We now show that the inequality on the right is fulfilled, namely, that $$\sum_{f \in \delta_{G_s}(\{u,v\})} x_f \leq 2.$$ Notice that whenever $y_v + y_u \leq 1$, the inequality is satisfied due to \eqref{eq:sumforuandv}, since the right-hand side would sum up to at most two and the left hand side only has nonnegative terms. 
    Now, assume that $y_v + y_u > 1$. Constraints \eqref{eq:sumforuandv} can be rewritten as
    \begin{equation}\label{eq:ineqright}
        \sum_{f \in \delta_{G_s}(\{u,v\})} x_f  = 2y_v + 2y_u - 2 x_e.
    \end{equation}
    We now claim that the right-hand side of \eqref{eq:ineqright} is at most two. Observe that \rafaelC{\eqref{cyc:05}}, together with the fact that we assumed that $y_v + y_u > 1$, implies that
    \begin{equation}
        2y_v + 2y_u - 2 x_e \leq 2y_v + 2y_u - 2(y_u + y_v -1) = 2.
    \end{equation}
     Observe that $x_e$ could be substituted by its lower bound as it was ensured to be positive, given the assumptions.
     
     As a consequence, $(\hat{x},\hat{y})$ satisfies \eqref{ger:04}.
\end{proof}

\begin{lemma}\label{lemma:bcwwy-cut-4}
    $Q^{cut} \subseteq Q^{BCWWy}$.
\end{lemma}
\begin{proof}
The proof follows from Lemmas~\ref{lemma:bcwwy-cut-1}-\ref{lemma:bcwwy-cut-3}.
\end{proof}

Now, consider $(\bar{x},\bar{y}) \in Q^{BCWWy}$. As Lemma~\ref{lemma:bcwwy-cut-1} already discusses the equivalence of several of the constraints, it only remains to show that $(\bar{x},\bar{y})$ satisfies constraints \eqref{cyc:04} and \eqref{cyc:05}.

\begin{lemma}\label{lemma:bcwwy-cut-5}
    Solution $(\bar{x},\bar{y})$ satisfies \eqref{cyc:04}.
\end{lemma}
\begin{proof}
    This is a consequence of constraints \eqref{ger:03} which imply the generalized subtour elimination constraints \eqref{subtour:01}. More specifically, observe that the latter imply for an edge $e=uv$, considering $S=\{u,v\}$, that $x_e \leq y_u$ and $x_e \leq y_v$. Consequently, $(\bar{x},\bar{y})$ satisfies \eqref{cyc:04}.
\end{proof}

\begin{lemma}\label{lemma:bcwwy-cut-6}
    Solution $(\bar{x},\bar{y})$ satisfies \eqref{cyc:05}.
\end{lemma}
\begin{proof}
    Remember that summing up inequalities \eqref{cyc:01} for $u$ and $v$, we obtain \eqref{eq:sumforuandv}, which can be rearranged as
    \begin{equation}\label{eq:sumforuandvrearranged}
         2 x_e = 2y_v + 2y_u - \sum_{f \in \delta_{G_s}(\{u,v\})} x_f.
    \end{equation}
    Note that \eqref{ger:04} imply that
    \begin{equation}
         2 x_e \geq 2y_v + 2y_u - 2,
    \end{equation}
    which dividing by two, guarantees that $(\bar{x},\bar{y})$ satisfies \eqref{cyc:05}.
\end{proof}

\begin{lemma}\label{lemma:bcwwy-cut-7}
    $Q^{BCWWy} \subseteq Q^{cut}$.
\end{lemma}
\begin{proof}
    This is a consequence of Lemmas \ref{lemma:bcwwy-cut-1}, \ref{lemma:bcwwy-cut-5}, and \ref{lemma:bcwwy-cut-6}.
\end{proof}

\begin{proposition}\label{proposition:bcwwy-cut-8}
    $Q^{BCWWy} = Q^{\rafaelC{cut}}$.
\end{proposition}
\begin{proof}
    It follows from Lemmas \ref{lemma:bcwwy-cut-4} and \ref{lemma:bcwwy-cut-7}.
\end{proof}

\begin{proposition}
    $Q^{BCWWy} \subset Q^{cec}$.
\end{proposition}
\begin{proof}
    It follows from Propositions \ref{proposition:cec-cut-4} and \ref{proposition:bcwwy-cut-8}.
\end{proof}

\newpage
\section{Comparing the strength of the different clique inequalities}
\label{sec:comparecliqueinequalities}

We claim that the clique inequality on the edge variables~\eqref{ineq:clique} is as strong as the inequality on the vertex variables~\eqref{ineq:cliquey}, under certain conditions. 


\begin{lemma}
\label{lemma:subcliquexy}
Given a clique $K \subseteq V$ of $G$, the inequality~\eqref{ineq:clique} implies the inequality~\eqref{ineq:cliquey} only if $$\sum\limits_{e \in \delta(K)} x_{e} \leq 2$$ is satisfied.
\end{lemma}
\begin{proof}
Consider a clique $K \subseteq V$ of $G$ and sum all the constraints~\eqref{cyc:02} corresponding to its vertices, which gives
$$\sum\limits_{e \in \delta(K)} x_{e} + \sum\limits_{e \in E(K)}2x_e = \sum\limits_{v \in K}2y_v.$$
Dividing by two in both sides we have that
$$\sum\limits_{e \in \delta(K)} \frac{1}{2}x_{e} + \sum\limits_{e \in E(K)}x_e = \sum\limits_{v \in K}y_v.$$
Assuming that \eqref{ineq:clique} is valid we have that
$$ \sum\limits_{v \in K}y_v \leq \sum\limits_{e \in \delta(K)} \frac{1}{2}x_{e} + 1.$$
Notice that, in order for $\sum\limits_{v \in K}y_v \leq 2$, one must have that $\sum\limits_{e \in \delta(K)} x_{e} \leq 2$. 
\end{proof}

\begin{lemma}
\label{lemma:subcliqueyx}
Given a clique $K \subseteq V$ of $G$, inequality~\eqref{ineq:cliquey} implies inequality~\eqref{ineq:clique} only if $$\sum\limits_{e \in \delta(K)} x_{e} \geq \sum\limits_{e \in E(K)} 2x_{e}$$ is satisfied.
\end{lemma}
\begin{proof}
Using the same reasoning employed in Lemma~\ref{lemma:subcliquexy}, we have that 
$$\sum\limits_{e \in \delta(K)} \frac{1}{2}x_{e} + \sum\limits_{e \in E(K)}x_e = \sum\limits_{v \in K}y_v.$$
Now, assuming that \eqref{ineq:cliquey} is valid, we obtain
$$\sum\limits_{e \in \delta(K)} \frac{1}{2}x_{e} + \sum\limits_{e \in E(K)}x_e \leq 2.$$
Notice that, in order to have $\sum\limits_{e \in E(K)}x_e \leq 1$ we must have that $\sum\limits_{e \in \delta(K)} \frac{1}{2} x_{e} \geq \sum\limits_{e \in E(K)} x_{e}$. Thus, the result follows. 
\end{proof}

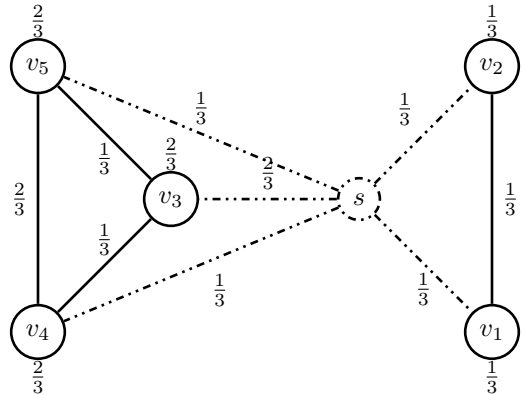
\begin{figure}[H]
	\centering
	\begin{tikzpicture}[auto, node distance=2.5cm]
	\tikzstyle{vertex}=[circle, draw,line width=1pt]
	\tikzstyle{vertex2}=[circle, draw,line width=1pt,dashdotdotted]
	\tikzstyle{inv}=[circle]
	\tikzstyle{edge} = [draw,thick,-,line width=1pt]
	\tikzstyle{edge2} = [draw,thick,-,line width=1pt,dashdotdotted]
	
	\node[vertex2] (s) {$s$};
	\node[inv] [above of = s, node distance=0.6cm] (sl) {};
	\node[vertex] [below right of = s] (v1) {$v_1$};
	\node[inv] [below of = v1, node distance=0.6cm] (l1) {$\frac{1}{3}$};
	\node[vertex] [above right of = s] (v2) {$v_2$};
	\node[inv] [above of = v2, node distance=0.6cm] (l2) {$\frac{1}{3}$};	
	\node[vertex] [left of = s] (v3) {$v_3$};
	\node[inv] [above of = v3, node distance=0.6cm] (l3) {$\frac{2}{3}$};	
	\node[vertex] [below left of = v3] (v4) {$v_4$};
	\node[inv] [below of = v4, node distance=0.6cm] (l4) {$\frac{2}{3}$};	
	\node[vertex] [above left of = v3] (v5) {$v_5$};
	\node[inv] [above of = v5, node distance=0.6cm] (l5) {$\frac{2}{3}$};	
	
	\path[edge2] (s) edge [below]node{$\frac{1}{3}$} (v1);        
	\path[edge2] (s) edge node{$\frac{1}{3}$} (v2);
	\path[edge2] (s) edge [above]node{$\frac{2}{3}$} (v3);
	\path[edge2] (s) edge node{$\frac{1}{3}$} (v4);
	\path[edge2] (s) edge [above]node{$\frac{1}{3}$} (v5);
	\path[edge] (v2) edge node{$\frac{1}{3}$} (v1);
	\path[edge] (v3) edge [above]node{$\frac{1}{3}$} (v4);
	\path[edge] (v4) edge node{$\frac{2}{3}$} (v5);
	\path[edge] (v5) edge [below]node{$\frac{1}{3}$} (v3);
	\end{tikzpicture}
	\caption{Example of feasible fractional solution that fulfills inequality~\eqref{ineq:cliquey}  but does not satisfy inequality~\eqref{ineq:clique}. Notice that the values corresponding to the edges linking the vertices in the clique $\{v_3,v_4,v_5\}$ sum up $\frac{4}{3}$, which is strictly greater than 1.}
	\label{fig:fractionalrespectvertex}
\end{figure}
\begin{figure}[H]
	\centering
	\begin{tikzpicture}[auto, node distance=2.5cm]
	\tikzstyle{vertex}=[circle, draw,line width=1pt]
	\tikzstyle{vertex2}=[circle, draw,line width=1pt,dashdotdotted]
	\tikzstyle{inv}=[circle]
	\tikzstyle{edge} = [draw,thick,-,line width=1pt]
	\tikzstyle{edge2} = [draw,thick,-,line width=1pt,dashdotdotted]
	
	\node[vertex2] (s) {$s$};
	\node[inv] [above of = s, node distance=0.6cm] (sl) {};	
	\node[vertex] [left of = s] (v3) {$v_3$};
	\node[inv] [above of = v3, node distance=0.6cm] (l3) {$\frac{4}{6}$};	
	\node[vertex] [below left of = v3] (v4) {$v_4$};
	\node[inv] [below of = v4, node distance=0.6cm] (l4) {$\frac{5}{6}$};	
	\node[vertex] [above left of = v3] (v5) {$v_5$};
	\node[inv] [above of = v5, node distance=0.6cm] (l5) {$\frac{4}{6}$};
	\node[vertex] [below left of = v4] (v1) {$v_1$};
	\node[inv] [below of = v1, node distance=0.6cm] (l1) {$\frac{1}{6}$};
	\node[vertex] [above left of = v4] (v2) {$v_2$};
	\node[inv] [above of = v2, node distance=0.6cm] (l2) {$\frac{1}{6}$};	
	
	\path[edge] (v4) edge [above]node{$\frac{1}{6}$} (v1);        
	\path[edge] (v4) edge [below]node{$\frac{1}{6}$} (v2);
	\path[edge2] (s) edge [above]node{$\frac{4}{6}$} (v3);
	\path[edge2] (s) edge [below]node{$\frac{4}{6}$} (v4);
	\path[edge2] (s) edge [above]node{$\frac{4}{6}$} (v5);
	\path[edge] (v2) edge [left]node{$\frac{1}{6}$} (v1);
	\path[edge] (v3) edge [above]node{$\frac{2}{6}$} (v4);
	\path[edge] (v4) edge node{$\frac{2}{6}$} (v5);
	\path[edge] (v5) edge [below]node{$\frac{2}{6}$} (v3);
	\end{tikzpicture}
	\caption{Example of feasible fractional solution that fulfills  inequality~\eqref{ineq:clique}  but does not satisfy inequality~\eqref{ineq:cliquey}. Notice that the values corresponding to the vertices in the clique $\{v_3,v_4,v_5\}$ sum up to $\frac{13}{4}$, which is strictly greater than 2.}
	\label{fig:fractionalrespectedge}
\end{figure}
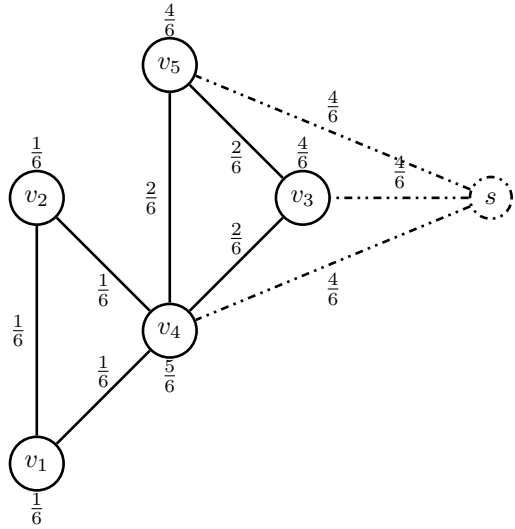

\begin{proposition}
In $Q^{cut}$, neither inequality~\eqref{ineq:cliquey} implies inequality~\eqref{ineq:clique} nor inequality~\eqref{ineq:clique} implies  inequality~\eqref{ineq:cliquey} 
\end{proposition}
\begin{proof}
The proof consists in showing two examples of feasible fractional solutions belonging to $Q^{cut}$. The first example, depicted in Figure~\ref{fig:fractionalrespectvertex}, presents a solution that fulfills the inequality~\eqref{ineq:cliquey}  but does not respect the inequality~\eqref{ineq:clique}. The second one, illustrated in Figure~\ref{fig:fractionalrespectedge}, presents a solution that satisfies the inequality~\eqref{ineq:clique} but does not respect the inequality~\eqref{ineq:cliquey}.

\end{proof}

\end{appendices}

\end{document}